\documentclass{du-journals}
\journal{Journal of Holography Applications in Physics}
\ArticleType{Regular article}
\Year{Spring 2025}
\Vol{5}
\No{2}
\Page{\thepage--\pageref{LastPage}}
\Received{January xx, 2025}
\Revised{January xx, 2025}
\Accepted{January xx, 2025}
\Doihead{10.22128/jhap.2021.452.xxxx}
\title{Holographic complexity and residual entropy of a rotating BTZ black hole within Horndeski gravity}
\subtitle{Holographic complexity and residual entropy of a rotating BTZ black hole within Horndeski gravity}
\author[1]{{Fabiano F. Santos}}
\address[1]{School of Physics, Damghan University, Damghan, 36716--41167, Iran. \\Departamento de Física, Universidade Federal do Maranhão, São Luís, 65080-805, Brazil.;\\ Corresponding Author E-mail: fabiano.ffs23@gmail.com}

\begin{document}

\begin{abstract}
This work explores the holographic complexity and residual entropy of a rotating BTZ black hole within the framework of Horndeski gravity. The investigation is motivated by the need to understand the emission of information from black holes, encoded by quantum complexity, which persists even at zero temperature. Traditionally, black holes are considered to cease emitting information upon reaching zero temperature, yet our findings suggest a minimum level of information or minimal entropy. This challenges the classical notion of black hole death. Recent studies in the context of Horndeski gravity and the AdS/BCFT correspondence have identified a nonzero minimal entropy at zero temperature. Our work shows that complexity and entropy provide crucial insights into the information emission from black holes, extending beyond their classical death. These findings significantly affect our understanding of black hole thermodynamics and quantum information theory.
\end{abstract}


\newpage

\tableofcontents

\newpage

\section{Introduction}

\begin{figure}[!ht]
\begin{center}
\includegraphics[width=\textwidth]{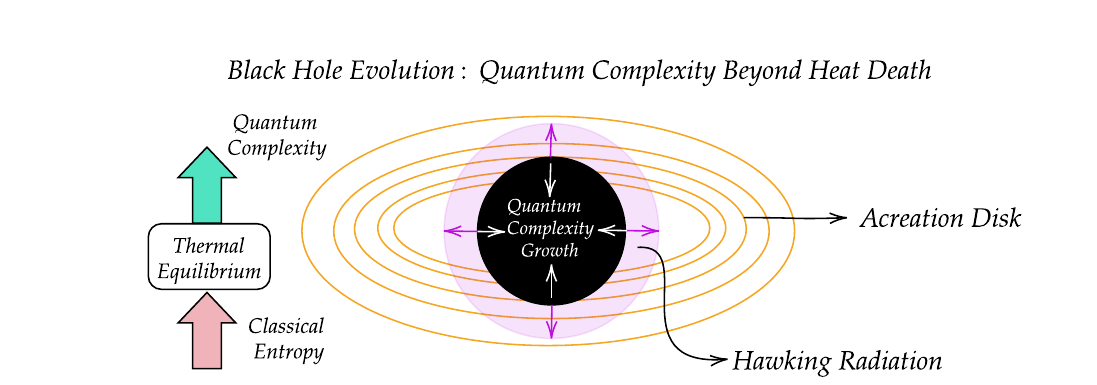}
\caption{The quantum information structure within black hole interiors exhibits continuous evolution through increasing state complexity, even as the black hole approaches thermodynamic equilibrium with its environment. This phenomenon suggests a form of persistent dynamical behavior beyond conventional thermodynamic heat death \cite{Susskind:2014moa}.}
\label{BHComplex}
\end{center}
\end{figure}

In recent years, the study of information processing in black holes has garnered significant interest (see {\sl e.g.}, \cite{Lloyd:2000cry}), particularly through the work of Leonard Susskind and collaborators \cite{Susskind:2014rva,Brown:2015bva,Brown:2015lvg,Susskind:2018fmx,Brown:2018bms}, which has led to the formulation of the second law of quantum complexity. This law suggests that black holes emit information encoded in quantum complexity \cite{Brown:2017jil}, challenging traditional views of black hole thermodynamics. Notably, the classical concept of a black hole's heat death, occurring when it reaches thermal equilibrium, is reinterpreted in this framework (see Fig. \ref{BHComplex}). These developments give an understanding of the black hole information dynamics and their implications for quantum gravity \cite{Susskind:2014moa,Raju:2020smc}.

Recent observations of supermassive black holes, particularly M87{*} \cite{Zhang:2024owe,Tsukamoto:2024gkz} and Sagittarius A* via the Event Horizon Telescope, provide indirect support for this theoretical framework by confirming the presence of complex accretion structures that align with predictions from quantum information models \cite{Burko:2023uag}. The implications extend to our understanding of the universe's ultimate fate, as these information-preserving processes may continue indefinitely, challenging traditional conceptions of cosmic heat death.

Susskind posits that complexity may underlie the phenomenon of black holes expanding beyond thermal equilibrium \cite{Susskind:2014rva,Susskind:2018fmx}. The vast number of quantum states in a system, which grows exponentially with qubits, implies that quantum computers and classical systems require significant time to explore the entire state space \cite{Brown:2019whu, Brown:2022rwi}. Consequently, the complexity of quantum states increases with time. Upon reaching thermal equilibrium, the interior size of a black hole could serve as a direct measure, enhancing our understanding of quantum black holes. The second law of quantum complexity, analogous to entropy, suggests that complexity increases until it reaches a maximum \cite{Brown:2017jil}. This law parallels the second law of classical thermodynamics and applies to black holes, though its broader implications for the universe remain uncertain. The concept of utilizing quantum circuit complexity offers a novel approach to studying quantum black holes \cite{Doroudiani:2019llj,Hashemi:2019aop}.

The preceding discussion suggests that black holes must emit information even at zero temperature in a quantum gravity framework. The anti-de Sitter/Conformal Field Theory (AdS/CFT) correspondence provides a powerful tool for exploring quantum gravity, allowing for a quantum description on the CFT side \cite{Maldacena:1997re, Witten:1998qj}. Recent studies in black hole thermodynamics have identified a non-zero minimal entropy at zero temperature, particularly within the context of Horndeski gravity \cite{Santos:2021orr, Sokoliuk:2022llp, Santos:2023flb, Santos:2023mee}. This minimal entropy is significant in scenarios where the AdS/BCFT correspondence is applied. For comprehensive reviews on Horndeski gravity, see \cite{Horndeski:1974wa,Horndeski:2024sjk, Charmousis:2011bf, Charmousis:2011ea, Bruneton:2012zk, Heisenberg:2018vsk, Kobayashi:2019hrl}. Furthermore, the evolution of holographic complexity in Horndeski gravity and beyond has been observed, aligning with predictions by Susskind and collaborators \cite{Susskind:2014rva, Brown:2015bva, Brown:2015lvg, Susskind:2018fmx, Brown:2018bms}, as discussed in \cite{Santos:2020xox, Santos:2022lxj, Bravo-Gaete:2020lzs}.

Studying the AdS\(_3\)/CFT\(_2\) correspondence within Horndeski gravity provides a way to explore how these modifications to gravity manifest in the dual CFT. For example, the scalar field in Horndeski gravity could correspond to specific operators in the CFT, and their interactions could reveal new insights about the nature of the duality \cite{Maldacena:1997re, Witten:1998qj}. Low dimensionality is one of the simplest and most mathematically tractable cases of the AdS/CFT correspondence. The symmetries of AdS\(_3\) (the isometry group \(SO(2,2)\)) match perfectly with the symmetries of a 2D CFT, which are governed by the infinite-dimensional Virasoro algebra \cite{Heisenberg:2018vsk}. This makes it a powerful testing ground for ideas in quantum gravity. In AdS\(_3\), black holes (e.g., BTZ black holes) are simpler to study, and their thermodynamic properties (like entropy) can be directly related to the CFT via the holographic renormalization \cite{Santos:2021orr, Sokoliuk:2022llp, Santos:2023flb, Santos:2023mee}. In Horndeski gravity, the presence of scalar fields can modify these black hole solutions, leading to new insights into how scalar-tensor theories affect holography.

In this work, we investigate the residual information of a rotating BTZ black hole within the Horndeski gravity framework, accessed through holographic complexity \cite{Banados:1992wn,Banados:1992gq}. This residual information, which persists at the Planck scale, is crucial for understanding the black hole's internal entropy or remnant \cite{Susskind:1995da}. Within the AdS/BCFT framework, we propose that these black holes possess an additional entropy component beyond the conventional Bekenstein-Hawking entropy. This minimal entropy, emerging from the black hole's heat death, is further influenced by boundary effects in the "Complexity=Action" (CA) approach, leading to corrections in holographic complexity \cite{Braccia:2019xxi, Aguilar-Gutierrez:2024rka, Zhou:2024pbb}. To quantify these effects, we analyze rotating BTZ black hole solutions in Horndeski gravity and derive the entropy using the AdS/BCFT correspondence as proposed by Takayanagi \cite{Takayanagi:2011zk}, with further insights from related works \cite{Fujita:2011fp, Fujita:2012fp, Melnikov:2012tb, Magan:2014dwa}.


Here we present  a  summary  of  the  main  results  achieved  in  this  work:

\begin{itemize}
\item Residual Information and Minimal Entropy: Using AdS/BCFT correspondence, we propose that these black holes possess a minimal entropy beyond the conventional Bekenstein-Hawking entropy. This challenges the classical notion of black hole heat death.
\item Holographic Complexity in Horndeski Gravity: The work explores the role of holographic complexity in understanding black hole thermodynamics, particularly through the AdS/BCFT correspondence. This minimal entropy, influenced by boundary effects in the "Complexity=Action" approach, introduces corrections to holographic complexity. 
\item Boundary Effects on Complexity: The study incorporates boundary effects in the "Complexity=Action" (CA) approach, leading to corrections in holographic complexity. 
\item Our analysis of rotating BTZ black hole solutions provides new insights into the interplay between quantum complexity, black hole thermodynamics, and the fundamental nature of quantum gravity.
\end{itemize}


This work is organized as follows. Section \ref{v1} presents our gravitational setup and how to combine it with BCFT theory.  In Section \ref{v2}, we consider a rotating BTZ black hole in Horndeski gravity and study the influence of the Horndeski parameter on the boundary Q profile (see Fig. \ref{AdSBCFT}). In Section \ref{v3}, we compute the entropy for a rotating BTZ black hole by performing a holographic renormalization. In Section \ref{CA12}, we present the corresponding holographic complexity and discuss the role played by the boundary in the "Complexity$=$Action" for the rotating BTZ black hole. Finally, Section \ref{v6} presents our conclusions and final comments.

\section{The Setup: AdS$_3$/BCFT$_2$ correspondence with Horndeski gravity}\label{v1}

In this section, we explore the configuration of a rotating BTZ black hole within the framework of AdS/BCFT duality and Horndeski gravity \cite{Banados:1992wn, Banados:1992gq, Santos:2020xox, Takayanagi:2011zk, Fujita:2011fp, Fujita:2012fp, Melnikov:2012tb, Magan:2014dwa}. Furthermore, the introduction of angular momentum follows the methodology outlined in \cite{Santos:2020xox}. 

\begin{figure}[!htbp]
  \includegraphics[width=\textwidth]{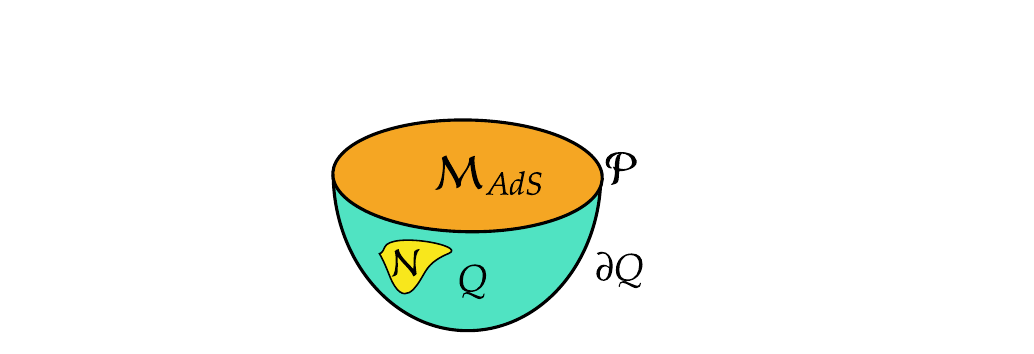}
    \caption{AdS/CFT correspondence in the presence of boundary hypersurface $Q$.}
    \label{AdSBCFT}
\end{figure}

Schematically, the AdS/BCFT duality consists in an extension of the AdS/CFT \cite{Maldacena:1997re} correspondence, defining in the CFT a boundary on the $d$-dimensional variety $ \mathcal{M}$ for an asymptotically $d+1$-dimensional AdS space $\mathcal{N}$ such that $\partial \mathcal{N}=\mathcal{M}~\cup~Q$, where $Q$ is a $d$-dimensional manifold satisfying $ \partial {Q}~\cap~\partial \mathcal{M}=\mathcal{P}$ (Fig. \ref{AdSBCFT}).

The contributions coming from surfaces $\mathcal{N}$, $Q$, and $\cal P$, in addition to the matter terms of $\mathcal{N}$ and $Q$ and the counterterms of $\cal P$ \cite{Takayanagi:2011zk, Fujita:2011fp, Fujita:2012fp, Melnikov:2012tb, Santos:2021orr, Magan:2014dwa}, provide the following action:
\begin{eqnarray}
S &=&\kappa\int_{\mathcal{N}}{d^{3}x\sqrt{-g}~\mathcal{L}_{\rm H}}+\kappa\int_{\mathcal{N}}{d^{3}x\sqrt{-g}~{\mathcal L}_{{\rm mat_1}}}
+2\kappa\int_{\rm bdry}{d^{2}x\sqrt{-h}~\mathcal{L}_{\rm bdry}} \nonumber\\
&&+2\int_{Q}{d^{2}x\sqrt{-h}~\mathcal{L}_{\rm mat_2}} +2\kappa\int_{\rm ct}{d^{2}x\sqrt{-h}~\mathcal{L}_{\rm ct}}\,, \label{S}
\end{eqnarray}
where ${\cal L}_{\rm H}$ is the Horndeski Lagrangian
\begin{eqnarray}
   {\cal L}_{\rm H} \equiv {\cal L}_{\rm EH} + {\cal L}_{\rm John} &=&(R-2\Lambda)-\frac{1}{2}(\alpha g_{\mu\nu}-\gamma\,  G_{\mu\nu})\nabla^{\mu}\phi\nabla^{\nu}\phi\,,
\end{eqnarray}
 which is defined here as the sum of the usual Einstein-Hilbert (EH) and the John sector Lagrangians \cite{Horndeski:1974wa,Horndeski:2024sjk, Charmousis:2011bf, Charmousis:2011ea, Bruneton:2012zk, Heisenberg:2018vsk, Kobayashi:2019hrl}. 
The matter terms, ${\cal L}_{\rm mat_1}$  and  $\mathcal{L}_{\rm mat_2}$ correspond to a perfect fluid in $\mathcal{N}$ and possible matter fields on $Q$, respectively. The boundary Gibbons-Hawking terms are given by \cite{Santos:2021orr}:
\begin{eqnarray}
\mathcal{L}_{\rm bdry}&=&(K-\Sigma)-\frac{\gamma}{4}(\nabla_{\mu}\phi\nabla_{\nu}\phi n^{\mu}n^{\nu}-(\nabla\phi)^{2})K-\frac{\gamma}{4}\nabla_{\mu}\phi\nabla_{\nu}\phi K^{\mu\nu}\,, \label{3}
\end{eqnarray}
With the $\gamma$-dependent contributions associated with Horndeski gravity.  $K=h^{\mu\nu}K_{\mu\nu}$, where $K_{\mu\nu}=h^{\beta}_{\mu}\nabla_{\beta}n_{\nu}$ is the extrinsic curvature, $h_{\mu\nu}$ the induced metric and $n^\mu$  the normal vector to the hypersurface $Q$ with tension $\Sigma$.  Furthermore, ${\cal L}_{\rm ct}$ are boundary counterterms 
\begin{eqnarray}
\mathcal{L}_{\rm ct} &=& c_{0}+c_{1}R+c_{2}R^{ij}R_{ij}+c_{3}R^{2}+b_{1}(\partial_{i}\phi\partial^{i}\phi)^{2}+...\,,\label{4}
\end{eqnarray}
localized on $\cal P$, which must be an asymptotic AdS spacetime.  

From the above definitions, the following total action follows.
\begin{eqnarray}
S&=&S^{\mathcal{N}}+S^{\mathcal{N}}_{\rm mat}+S^{Q}+S^{Q}_{mat}+S^{P}_{ct}\,\label{S}
\end{eqnarray}
and imposing a Neumann boundary condition, we have 
\begin{eqnarray}
K_{\alpha\beta}-h_{\alpha\beta}(K-\Sigma)-\frac{\gamma}{4}H_{\alpha\beta}=\kappa \, {\cal S}^{Q}_{\alpha\beta}\,,\label{5}
\end{eqnarray}
where we defined 
\begin{eqnarray}
&&{H_{\alpha\beta}\equiv(\nabla_{\mu}\phi\nabla_{\nu}\phi \, n^{\mu}n^{\nu}-(\nabla\phi)^{2})(K_{\alpha\beta}-h_{\alpha\beta}K)-(\nabla_{\mu}\phi\nabla^{\mu}\phi)h_{\alpha\beta}K}\,,\label{6}\\
&&{\cal S}^{Q}_{\alpha\beta}=-\frac{2}{\sqrt{-h}}\frac{\delta S^{Q}_{mat}}{\delta h^{\alpha\beta}}\,.\label{7} 
\end{eqnarray}

Considering $S^{Q}_{mat}$ as constant, we have ${\cal S}^{Q}_{\alpha\beta}=0$. Then, we can write
\begin{eqnarray}
K_{\alpha\beta}-h_{\alpha\beta}(K-\Sigma)-\frac{\gamma}{4}H_{\alpha\beta}=0\,.\label{8}
\end{eqnarray}
Assuming that in the bulk $S^{\mathcal{N}}_{\rm mat}$ is also a constant and varying $S^{\mathcal{N}}$ with respect to $g_{\alpha\beta}$ and $\phi$, and $S^Q$ with respect to $\phi$, we have the following.
\begin{eqnarray}
{\cal E}_{\alpha\beta}[g_{\mu\nu},\phi]=-\frac{2}{\sqrt{-g}}\frac{\delta S^{\mathcal{N}}}{\delta g^{\alpha\beta}}\,,\quad {\cal E}_{\phi}[g_{\mu\nu},\phi]=-\frac{2}{\sqrt{-g}}\frac{\delta S^{\mathcal{N}}}{\delta\phi} \,,\quad {\cal F}_{\phi}[g_{\mu\nu},\phi]=-\frac{2}{\sqrt{-h}}\frac{\delta S^{Q}}{\delta\phi} \,.\nonumber\\
\end{eqnarray}
Then, one finds: 
\begin{eqnarray}
{\cal E}_{\mu\nu}[g_{\mu\nu},\phi]&=&G_{\mu\nu}+\Lambda g_{\mu\nu}-\frac{\alpha}{2}\left(\nabla_{\mu}\phi\nabla_{\nu}\phi-\frac{1}{2}g_{\mu\nu}\nabla_{\lambda}\phi\nabla^{\lambda}\phi\right)\label{11}\nonumber\\
                  &+&\frac{\gamma}{2}\left(\frac{1}{2}\nabla_{\mu}\phi\nabla_{\nu}\phi R-2\nabla_{\lambda}\phi\nabla_{(\mu}\phi R^{\lambda}_{\nu)}-\nabla^{\lambda}\phi\nabla^{\rho}\phi R_{\mu\lambda\nu\rho}\right)\nonumber\\
									&+&\frac{\gamma}{2}\left(-(\nabla_{\mu}\nabla^{\lambda}\phi)(\nabla_{\nu}\nabla_{\lambda}\phi)+(\nabla_{\mu}\nabla_{\nu}\phi)\Box\phi+\frac{1}{2}G_{\mu\nu}(\nabla\phi)^{2}\right)\nonumber\\
									&-&\frac{\gamma\,g_{\mu\nu}}{2}\left(-\frac{1}{2}(\nabla^{\lambda}\nabla^{\rho}\phi)(\nabla_{\lambda}\nabla_{\rho}\phi)+\frac{1}{2}(\Box\phi)^{2}-(\nabla_{\lambda}\phi\nabla_{\rho}\phi)R^{\lambda\rho}\right),\\
{\cal E}_{\phi}[g_{\mu\nu},\phi]&=&\nabla_{\mu}[(\alpha g^{\mu\nu}-\gamma\,G^{\mu\nu})\nabla_{\nu}\phi]\,,\label{12}\\
{\cal F}_{\phi}[g_{\mu\nu},\phi]&=&-\frac{\gamma}{4}(\nabla_{\mu}\nabla_{\nu}\phi n^{\mu}n^{\nu}-(\nabla^{2}\phi))K-\frac{\gamma}{4}(\nabla_{\mu}\nabla_{\nu}\phi)K^{\mu\nu}\,,\label{12.1}
\end{eqnarray}
Furthermore, the Euler-Lagrange equations imply that ${\cal E}_{\phi}[g_{\mu\nu},\phi]={\cal F}_{\phi}[g_{\mu\nu},\phi]$.

\section{Q-profile within rotating BTZ black hole in Horndeski gravity}\label{v2}

In this section, we describe our BTZ black hole and construct the profile of the hypersurface $Q$, considering the influence of Horndeski gravity. We will introduce the angular momentum $J$ for the BTZ black hole, using the metric \cite{Santos:2020xox}:
\begin{eqnarray}
ds^{2}=-f(r)dt^{2}+r^{2}\left(dy-\frac{J}{r^{2}}dt\right)^{2}+\frac{dr^{2}}{f(r)}\,.
\label{13}
\end{eqnarray}
To address the static configurations of black holes, certain Galileon models allow for spherically symmetric solutions, as demonstrated by \cite{Bravo-Gaete:2013dca} in the context of the no-hair theorem. This theorem stipulates that the squared radial component of the conserved current must vanish identically without imposing constraints on the radial dependence of the scalar field. This condition implies:
$$ \alpha g_{rr} -\gamma\,G_{rr}=0 \,.$$
Combining this condition with ${\cal E}_{\phi}[g_{rr},\phi]={\cal E}_{rr}[g_{rr},\phi]=0$, we have
\begin{eqnarray}
f(r)&=&-M^2+\frac{\alpha r^{2}}{\gamma}+\frac{J^{2}}{r^{2}},\label{15}\\
\psi^{2}(r)&=&-\frac{2(\alpha+\gamma\Lambda)}{\alpha\gamma\,f(r)}.\label{16}
\end{eqnarray}
\noindent 
These solutions can be asymptotically dS or AdS for $\alpha/\gamma<0$ or $\alpha/\gamma>0$, respectively. The structure of the black hole's horizon can be studied using equation (\ref{15}), for which there are two roots for the function $f(r)$, i.e.,  
\begin{eqnarray}\label{r+-}
r_{\pm}=\sqrt{\frac{\gamma\,M^2}{2\alpha}\pm\frac{\gamma}{2\alpha}\sqrt{M^4-\frac{4\gamma\,J^2}{\alpha}}}. 
\end{eqnarray}

\begin{figure}[!ht]

\begin{center}

\includegraphics[width=\textwidth]{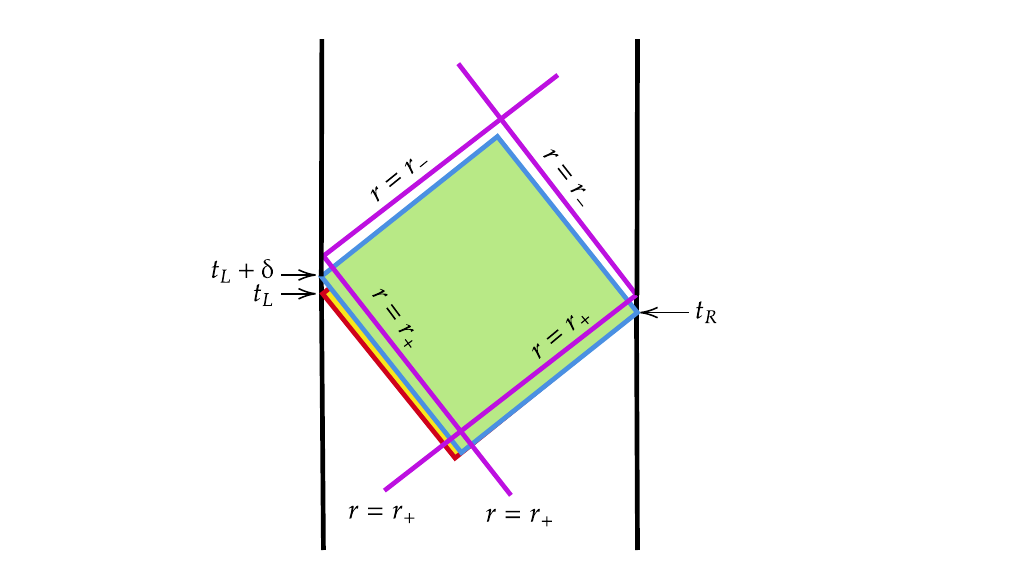}
\caption{This figure shows that the Wheeler-DeWitt patch for the rotating BTZ black hole does not extend to the singularity ending in incoming light sheets intersecting outside the inner horizon at $r_{-}$.}
\label{pensr}
\label{ylinhaz}
\end{center}
\end{figure}

In its formation process, when we add angular momentum to an AdS black hole, we change how the Wheeler-DeWitt patch ends \cite{Brown:2015lvg}. Thus, instead of ending when the incoming light sheets collide with the singularity at $r=0$, we can observe according to Fig. \ref{pensr} that light sheets collide with each other at $t=0$ (for $t_{L}=t_{R}$). However, we can follow \cite{Santos:2020xox} to find the late growth of the complexity.

In the formation process of an AdS black hole, the addition of angular momentum alters the termination of the Wheeler-DeWitt patch \cite{Brown:2015lvg}. Typically, this patch ends when incoming light sheets intersect the singularity at $r=0$. However, as illustrated in Fig. \ref{pensr}, with angular momentum, these light sheets instead collide at  $t=0$ (for $t_{L}=t_{R}$). This modification has significant implications for the late-time growth of complexity, which can be further explored following the methodology outlined in \cite{Santos:2020xox}.

In Fig. \ref{pensr}, the entire WDW patch lies inside the inner horizon at $r_{+}$. If $r_{L}$ increases, the Wheeler-DeWitt patch gains a slice (in blue) and loses a slice (in red). The idea behind these discussions shows us that the action is not sensitive to the quantum instabilities of the internal horizon as long as the horizon remains null. On the other hand, classical instabilities are not considered here, as they lead to significant changes in the structure of the internal horizon \cite{Brown:2015lvg}.

 Therefore, the area of interest for our studies is precisely $r_{+}$, as it is precisely where the WDW patch is contained; see Fig. \ref{pensr}. However, as discussed in previous work \cite{Santos:2021orr}, one can perform a rescaling in $r_\pm$, Eq. \eqref{r+-}, and rewrite 
\begin{eqnarray}
r^{2}_{\pm}=\frac{M^2}{2}\pm\frac{1}{2}\sqrt{M^4-4J^2}. 
\end{eqnarray}

\begin{figure}[!ht]
\begin{center}
\includegraphics[width=\textwidth]{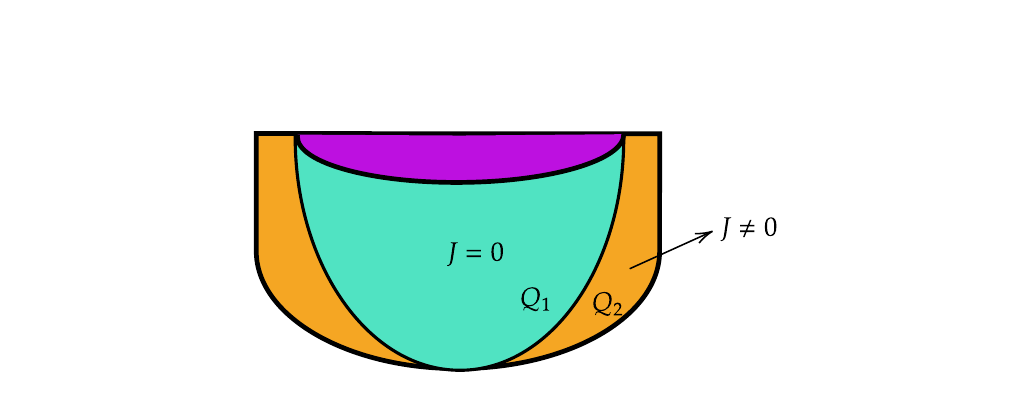}
\caption{This figure shows how the angular momentum affects the rotating BTZ black hole, represented by boundary $Q_2$. Note that, in the case where $J=0$, we have $Q_1=Q_2$.}
\label{BTZ1}
\end{center}
\end{figure}

\begin{figure}[!ht]
\begin{center}
\includegraphics[scale=0.75]{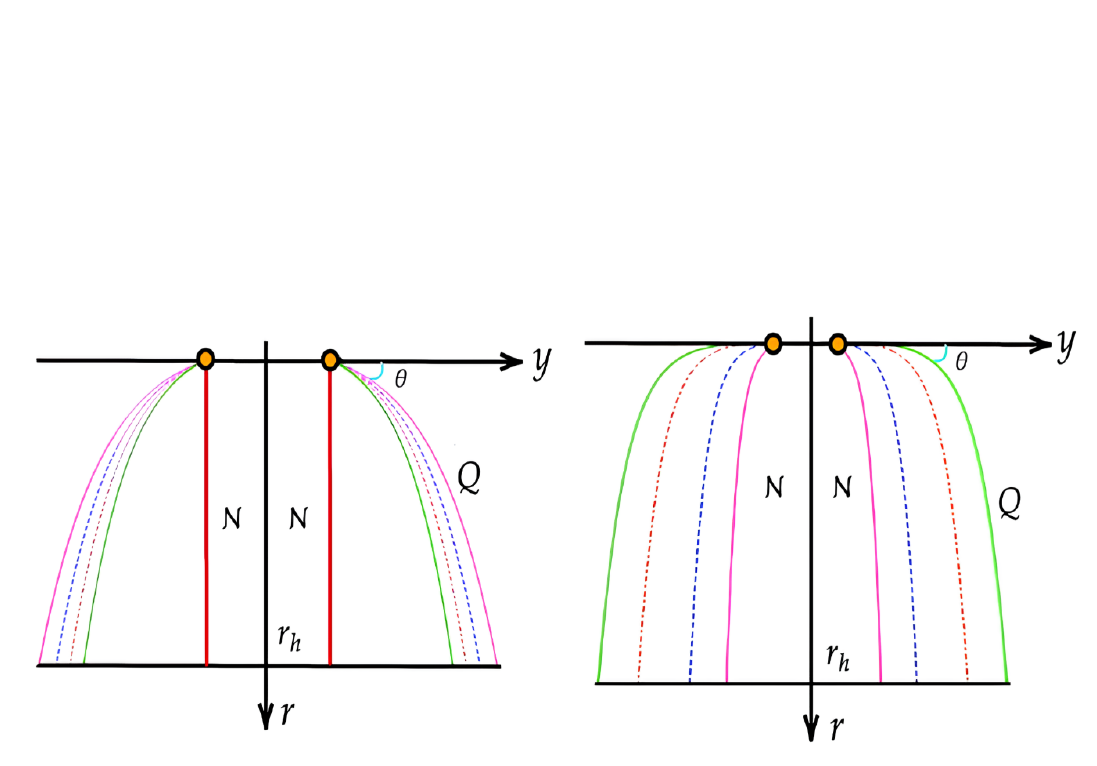}
\caption{{\sl Left panel:} Q boundary profile for the BTZ black hole within Horndeski gravity considering the values for $J=0$, $M=-1$, $\theta'=2\pi/3$, $\theta=\pi-\theta'$,  $\alpha=-8/3$ with $\gamma=-0.1$ ({\sl solid}), $\gamma=-0.2$ ({\sl dashed}), $\gamma=-0.3$ ({\sl dot dashed}), and $\gamma=-0.4$ ({\sl thick}). The dashed parallel vertical lines represent the UV solution, Eq. \eqref{19}. The region between curve Q's negative and positive branches represents the bulk $N$ \cite{Santos:2021orr}. {\sl Right panel:} present the same values of the {\sl left panel:} for the Horndeski parameters; now, we have with $M=1.1$ and $J=0.2$ ({\sl solid}), $M=1.2$ and $J=0.4$ ({\sl dashed}), $M=1.4$ and $J=0.6$ ({\sl dot dashed}), $M=1.6$ and $J=0.8$ ({\sl thick}).}
\label{profile}
\end{center}
\end{figure}

Now, to construct the $Q$ boundary profile, one has to use the induced metric on this surface given by 
\begin{eqnarray}
ds^{2}_{\rm ind}=-\left(f(r)-\frac{J^2}{r^2}\right)dt^{2}+\frac{g^{2}(r)dr^{2}}{f(r)}-2Jdydt\,, 
\end{eqnarray}
where $g^{2}(r)=1+r^2{y'}^{2}(r)f(r)$ with $y{'}(r)=dy/dr$. Then, the normal vectors on $Q$ can be represented by 
\begin{eqnarray}
n^{\mu}=\frac{1}{g(r)}\, \left(0,\, 1, \, -{f(r)y{'}(r)}\right)\,.\label{17}
\end{eqnarray}
Fulfilling the no-hair theorem, meaning ${\cal F}_{\phi}[h_{rr},\phi]=0$, one can solve the Eq. \eqref{8}, so that 
\begin{eqnarray}
y{'}(r)&=&\frac{(\Sigma l_{AdS})}{\sqrt{1+\dfrac{\gamma\psi^{2}(r)}{4}-(\Sigma l_{AdS})^{2}f(r)}}\,, 
\cr \cr 
&=&\frac{(\Sigma l_{AdS})}{\sqrt{1-\dfrac{\xi}{f(r)}-(\Sigma l_{AdS})^{2}f(r)}}\,,
\label{19}
\end{eqnarray}
\noindent with $\psi(r)$ given by Eq. \eqref{16}, so that $\xi$ is defined as 
\begin{equation}
     \xi=-\frac{1}{2}\frac{\alpha+\gamma\Lambda}{\alpha}
     \,. 
\end{equation}
The profiles implied by these solutions are shown in Figs. \ref{BTZ1} and \ref{profile}. \\

For zero angular momentum, the boundary region $Q_1$ (see Fig. \ref{BTZ1}) at finite temperature serves as a candidate for the bulk geometry \cite{Santos:2021orr}. The AdS/BCFT correspondence is applied to this configuration as illustrated in Fig. \ref{profile}. As discussed in Section \ref{v3}, the residual entropy information becomes significant in the low temperature regime. Introducing fixed angular momentum results in a widening of the boundary \cite{Long:2020wqj}. In this context, the "Randall-Sundrum brane" AdS$_2$ within AdS$_3$ is deformed by the angular momentum of the BTZ black hole \cite{Nozaki:2012qd}. This deformation corresponds to the boundary $Q_2$ in the region where $y_{UV}(r)=y_{0}$, which is perpendicular to $\mathcal{M}_{AdS_{3}}$.

The profile behavior shown in Fig. \ref{profile} is computed using a numerical procedure. Replacing the holographic renormalization procedure to study the residual information through the total entropy is difficult. However, since $\xi$ is a small parameter, we can expand Eq. (\ref{19}) around this value up to first order:
\begin{eqnarray}
y_{Q_{2}}\equiv y(r)&=&y_{0}+\int{\frac{(\Sigma l_{AdS})dr}{\sqrt{1-(\Sigma l_{AdS})^{2}f(r)}}}+\int{\frac{\xi(\Sigma l_{AdS})dr}{2f(r)[1-(\Sigma l_{AdS})^{2}f(r)]^{3/2}}}+\mathcal{O}(\xi)\,. \label{19.3}
\end{eqnarray}
For a rotating BTZ black hole, the Wheeler-DeWitt patch does not extend to the singularity; it ends when the incoming sheets of light intersect outside the inner horizon at $r_{-}$. Thus, we can evaluate the integrals Eq. (\ref{19.3}) in the Wheeler-DeWitt patch region; see Fig. \ref{pensr}. 

\section{Black hole entropy}\label{v3} 
In this section, we present a holographic scheme with angular momentum contributions for the AdS/BCFT correspondence within Horndeski gravity. 

\begin{figure}[!ht]
\begin{center}
\includegraphics[width=\textwidth]{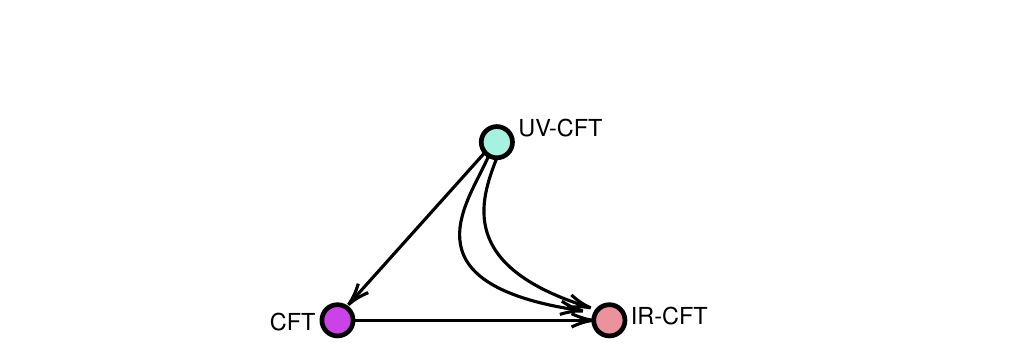}
\caption{Organized scheme of CFT space.}
\label{BTZ2}
\end{center}
\end{figure}

Let us then start with the Euclidean action given by
\begin{equation}\label{IE}
I_{\rm E}=I_{\rm bulk}
+2I_{\rm bdry},
\end{equation}
where the bulk Euclidean action is 
\begin{eqnarray}
&&I_{\rm bulk}=-\frac{1}{2} \kappa\int_{\mathcal{N}}{d^{3}x\sqrt{g}\left[(R-2\Lambda)+\frac{\gamma}{2} G_{\mu\nu}\nabla^{\mu}\phi\nabla^{\nu}\phi\right]}\cr 
&&-\frac{1}{2}\int_{\mathcal{M}}{d^{2}x\sqrt{\bar{\gamma}}\left[(K^{(\bar{\gamma})}-\Sigma^{(\bar{\gamma})})-\frac{\gamma}{4}(\nabla_{\mu}\phi\nabla_{\nu}\phi n^{\mu}n^{\nu}-(\nabla\phi)^{2})K^{(\bar{\gamma})}-\frac{\gamma}{4}\nabla^{\mu}\phi\nabla^{\nu}\phi K^{(\bar{\gamma})}_{\mu\nu}\right]},\cr
 &&\label{BT}
\end{eqnarray}
with $\kappa^{-1}={8\pi G_{N}}$ being the gravitational coupling, $g$ the determinant of the metric $g_{\mu\nu}$ on the bulk $\mathcal{N}$, $\bar{\gamma}$ the induced metric on the surface $\cal M$ with tension tension  $\Sigma^{(\bar{\gamma})}$, and  extrinsic curvature with trace  $K^{(\bar{\gamma})}$.
 On the other hand, for the boundary, one has the Euclidean action 
\begin{eqnarray}
I_{\rm bdry}&=&-\frac{1}{2} \kappa \int_{\mathcal{N}}{d^{3}x\sqrt{g}\left[(R-2\Lambda)
+\frac{\gamma}{2}G_{\mu\nu}\nabla^{\mu}\phi\nabla^{\nu}\phi\right]}\cr
&&-\kappa \int_{Q}{d^{2}x\sqrt{h}\left[(K-\Sigma)-\frac{\gamma}{4}(\nabla_{\mu}\phi\nabla_{\nu}\phi n^{\mu}n^{\nu}-(\nabla\phi)^{2})K-\frac{\gamma}{4}\nabla^{\mu}\phi\nabla^{\nu}\phi K_{\mu\nu}\right]}.\cr
&&\label{BT1}
\end{eqnarray}
The AdS/CFT correspondence shows that IR divergences on the gravity side correspond to the UV divergences on the CFT boundary theory. This relation is the IR-UV connection; see Fig. \ref{BTZ2}. 

\begin{figure}[!ht]
\begin{center}
\includegraphics[width=\textwidth]{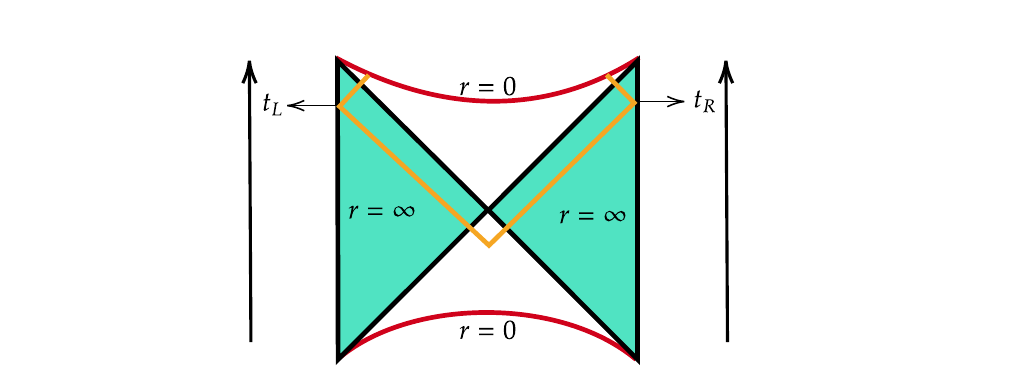}
\caption{This diagram presents the evolution of complexity as the entropy of the BTZ black hole increases. In this configuration, the bilateral AdS black hole is dual to an entangled state (dual thermal field) of two CFTs living at the left and right boundaries. From the complexity/action point of view, the complexity of the CFT state is the same as the action of the Wheeler-DeWitt patch (violet lines inside the Penrose diagram).}
\label{BTZ3}
\end{center}
\end{figure}

The Bekenstein-Hawking (BH) entropy \cite{Banerjee:2019vff} that can be found through the free energy is defined as
\begin{equation}\label{FE}
 \Omega=T_{\rm H}\, I_{\rm E} \,,  
\end{equation}
where $T_{\rm H}$ is the Hawking temperature. Therefore, one can obtain the corresponding entropy as:
\begin{eqnarray}
S_{\rm BH}=-\frac{\partial\Omega}{\partial T_{\rm H}} = - I_E\,.\label{BT7}
\end{eqnarray}
Using the results from the Euclidean actions, Eqs. \eqref{IE}, \eqref{BT}, and \eqref{BT1}, together with the solutions, Eq. \eqref{19.3}, 
one finds for small $\xi$ that 
\begin{eqnarray}
&&S_{\rm BH}=\frac{\Delta y}{4G_{N}}\left(1-\frac{\xi}{4}\right)\sqrt{\frac{M^2}{2}+\frac{1}{2}\sqrt{M^4-4J^2}}-\frac{\Delta y_{Q_{2}}}{G_{N}}\left(1-\frac{\xi}{8}\right)\sqrt{\frac{M^2}{2}+\frac{1}{2}\sqrt{M^4-4J^2}}\cr
&&-\xi\,l^3_{AdS}\,p(\theta{'})\Delta\,y_{Q_{2}}\left[\frac{M^2}{2}+\frac{1}{2}\sqrt{M^4-4J^2}\right]+\frac{\xi\,l_{AdS}q(\theta{'})\Delta y_{Q_{2}}}{4G_{N}}
\end{eqnarray}
where
\begin{eqnarray}
&&p(\theta{'})=2b(\theta{'})\left(1-\frac{\xi}{8}\right)+6h(\theta{'})\cot(\theta{'}),\\
&&b(\theta{'})=\cos(\theta{'})\tan^{-1}\left(\frac{1}{\sin(\theta{'})}\right)+\cot(\theta{'})\left(\frac{1+\cos^{2}(\theta{'})\cot^{2}(\theta{'})}{\sin^{2}(\theta{'})}\right),\\
&&h(\theta{'})=-\frac{(1+\pi/2)}{2\sin(\theta{'})}+\frac{\cot^{3}(\theta{'})\cos^{2}(\theta{'})}{(1+\cos^{2}(\theta{'}))}\tanh^{-1}\left(\frac{\sqrt{2}\cos(\theta{'})}{\sqrt{1+\cos^{2}(\theta{'})}}\right).
\end{eqnarray}
Then, one can see that this entropy contains information about the black hole: mass and angular momentum. This result can be compared with Ref \cite{Banerjee:2019vff} for rotating black holes in AdS space. If no charge can be radiated as the Hawking black hole evaporates completely, we have a BTZ black hole with \( M = 0 \) and \( J = 0 \), which corresponds to
\begin{eqnarray}
S^{\rm residual}_{\rm BH} =\frac{\xi\,l_{AdS}q(\theta{'})\Delta y_{Q_{2}}}{4G_{N}}
\end{eqnarray}
This quantity characterizes the residual information associated with boundary observers over a finite time or bulk observers lacking access to certain spacetime regions within the Wheeler-DeWitt (WDW) patch. These findings support the assertion by Brown et al. \cite{Brown:2017jil} that black hole heat death occurs only classically. At the same time, the information content continues to increase due to quantum complexity, which follows a second law for black holes (see Fig. \ref{BTZ3}). Consequently, we propose that minimal entropy $S^{residual}_{BH}$ could represent a boundary Conformal Field Theory (BCFT) "fundamental state".

In simple terms, residual entropy refers to the idea that even after a black hole has radiated most of its energy through Hawking radiation, it does not completely disappear \cite{Hosseini:2017mds,Cabo-Bizet:2018ehj,Choi:2018hmj}. Instead, a small amount of entropy, or "information," remains. This leftover entropy is tied to the black hole's internal structure and quantum properties, which persist at the Planck scale (the smallest scale of the universe where quantum gravity effects dominate). Traditionally, black holes were thought to evaporate entirely, leaving behind no trace. However, our residual entropy, including those using the AdS/BCFT (Anti-de Sitter/Boundary Conformal Field Theory) framework, suggests that a minimal entropy remains even at zero temperature  \cite{Santos:2021orr}. This challenges the classical notion of black hole "death" and implies that black holes retain a remnant—a tiny, stable object that holds the remaining information. This residual entropy is significant because it provides a way to reconcile black hole thermodynamics with quantum mechanics. It suggests that information falling into a black hole is not lost forever but is encoded in the remnant. This aligns with the idea that black holes are not just destructive entities but also play a role in preserving the fundamental information of the universe.

\section{Holographic Complexity}\label{CA12}

To study the growth of the holographic complexity of the rotating BTZ black hole, according to \cite{Santos:2020xox}, considering the parameters $\tau$ and $\sigma$ in the world-sheet of the fundamental string. These parameters are given as follows:
\begin{eqnarray}
t=\tau,\quad r=\sigma\quad,y=v\tau+\zeta(\sigma).
\end{eqnarray}
We can write for the induced metric in the bulk:
\begin{eqnarray}
\frac{1}{T_{s}}\frac{dS^{\rm bulk}_{NG}}{d(t_L+t_R)}=\frac{1}{T_{s}}\frac{dS^{\rm bulk}_{NG}}{dt}=\left(1-\frac{\xi}{4}\right)\sqrt{\left[\frac{M^2}{2}+\frac{1}{2}\sqrt{M^4-4J^2}\right]}\label{complexiy}
\end{eqnarray}
 Now, we need to compute the boundary contributions for “$Complexity=Action$” (CA) 
\begin{eqnarray}
\frac{1}{T_{s}}\frac{dS^{\rm boundary}_{NG}}{d(t_L+t_R)}=\frac{1}{T_{s}}\frac{dS^{\rm boundary}_{NG}}{dt}=\left(1-\frac{\xi}{4}\right)\int^{r_{+}}_{0}{d\sigma\zeta{'}(\sigma)}=\left(1-\frac{\xi}{4}\right)\zeta(\sigma)|_{\sigma=r_{+}}\label{complexiy1}
\end{eqnarray}
where through the parameters $\tau$ and $\sigma$ in the world-sheet, we can see that $\zeta(\sigma)=y(\sigma)$, i.e., the embedding of the string $\zeta(\sigma)$ is described by the boundary profile: 
\begin{eqnarray}
\zeta(\sigma)=\zeta_{0}+\int{\frac{d\sigma(\Sigma l_{AdS})}{\sqrt{1-\dfrac{\xi}{f(\sigma)}-(\Sigma l_{AdS})^{2}f(\sigma)}}}\label{complexiy2}
\end{eqnarray}

Now, we study the UV and IR regimes. Considering the IR case and performing an expansion with $\sigma\to\infty$, Eq. (\ref{complexiy2}) becomes 

\begin{eqnarray}
\zeta_{_{IR}}(\sigma)=\zeta_{0}+\frac{\ln(\sigma)}{\sqrt{-\alpha/\gamma}}.\label{19.2}
\end{eqnarray}
where the complexity becomes
\begin{eqnarray}
\frac{1}{T_{s}}\frac{dS^{\rm boundary}_{NG}}{dt}=\frac{1}{\sqrt{-\alpha/\gamma}}\ln\left(\sqrt{\left[\frac{M^2}{2}+\frac{1}{2}\sqrt{M^4-4J^2}\right]}\right)\label{complexiy3}
\end{eqnarray}
On the other hand, for the UV regime ($\sigma\to\,0$), we have 
\begin{eqnarray}
\zeta_{_{IR}}(\sigma)\sim\zeta_{0}+\left(\frac{M^2}{2}+\frac{1}{2}\sqrt{M^4-4J^2}\right).\label{zeta1}
\end{eqnarray}
where the Eq. (\ref{zeta1}) provides the following equation for the complexity
\begin{eqnarray}
\frac{1}{T_{s}}\frac{dS^{\rm boundary}_{NG}}{dt}=\frac{1}{2}\left(M^2+\sqrt{M^4-4J^2}\right)\label{complexiy4}
\end{eqnarray}
The equation above indicates that for rotating BTZ black holes, the ground state corresponds to $M$ when $J=0$. Consequently, the rate of change of action in the Wheeler-DeWitt (WDW) patch for a BTZ black hole reaches saturation \cite{Brown:2015bva}. We observe that the "$Complexity=Action$" boundary data primarily affects the finite term of the expansion as the ultraviolet (UV) cutoff is removed. Our analysis confirms that, in the case of a free boson, this divergence is inherently a boundary feature and is therefore absent \cite{Braccia:2019xxi}.

The rotating BTZ black hole is incorporated into a complete ultraviolet (UV) theory without developing hair, in agreement with the weak gravity conjecture \cite{Arkani-Hamed:2006emk}. However, an apparent violation of the complexity threshold could serve as an indicator of hair development. By expanding Eq. (\ref{complexiy2}) around $\xi$ to first order, we find that the total complexity includes contributions from both the bulk UV and infrared (IR) regions:
\begin{eqnarray}
\frac{1}{T_ {s}}\frac{dS^{\rm total}_{NG}}{dt}
&=&\left(1-\frac{\xi}{4}\right)\sqrt{\left[\frac{M^2}{2}+\frac{1}{2}\sqrt{M^4-4J^2}\right]}+\frac{\Delta\zeta^{\rm UV}_{Q_2}}{2} \left(1-\frac{\xi}{4}\right)\left(M^2+\sqrt{M^4-4J^2}\right)\cr
&& +\frac{\Delta\zeta^{\rm IR}_{Q_2}}{2\sqrt{-\alpha/\gamma}} \left(1-\frac{\xi}{4}\right)\ln(M^2+\sqrt{M^4-4J^2})-\frac{\Delta\zeta^{\rm IR}_{Q_2}}{2\sqrt{-\alpha/\gamma}} \left(1-\frac{\xi}{4}\right)\ln(2)\label{complexiy6}
\end{eqnarray}

In our framework, the $Complexity=Action$ (CA) conjecture identifies states that do not conform with a consistent truncation of a UV-complete theory \cite{Zhou:2024pbb}. The Wheeler-DeWitt (WDW) patch emerges as the natural spacetime region associated with a limit state, demonstrating robustness against small perturbations. Consequently, the complete Horndeski action and the BCFT boundary component become intrinsically linked to the WDW patch. This relationship has been extended to higher-dimensional generalizations beyond the BTZ black hole \cite{Susskind:2014rva, Brown:2015bva, Brown:2015lvg, Brown:2017jil}. Notably, as discussed by Brown et al. \cite{Brown:2017jil}, neutral AdS black holes, regardless of dimensionality or size, saturate the same limit at the computation rate with a consistent coefficient.

\begin{figure}[!ht]
\begin{center}
\includegraphics[width=\textwidth]{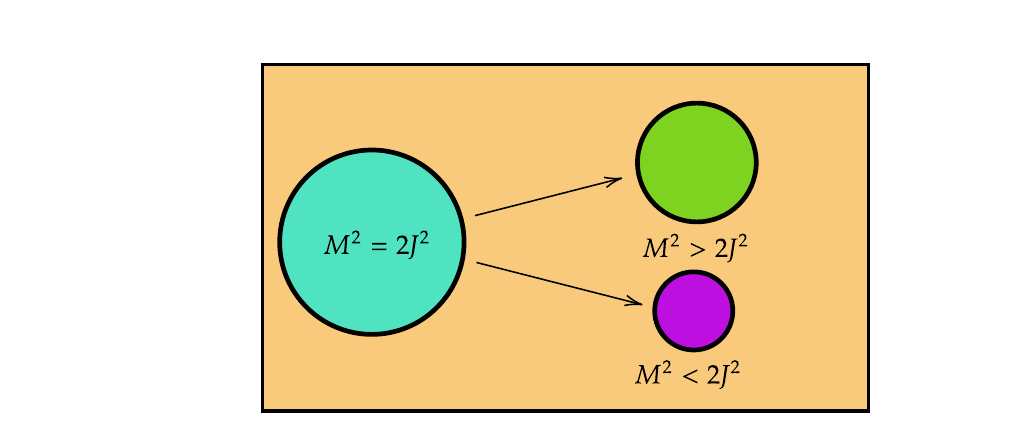}
\caption{This figure presents some possibilities for the mass $M$ and $J$, which involve the black hole's angular momentum. Note that an extremal black hole ($M^2=2J^2$) can only decay if there are particles whose charge exceeds its mass.}
\label{BTZ4}
\end{center}
\end{figure}

The solution for the rotating BTZ black hole in the regime $M^2\geq\,2J^2$  indicates that no charge is radiated in the absence of very light-charged particles as the black hole evaporates to the Planck scale. At this scale, the effects of Horndeski gravity and quantum mechanics become significant. Our approach, equipped with the AdS/BCFT correspondence, provides a satisfactory quantum description of gravity via Conformal Field Theory with boundaries (BCFT's). However, Planck-scale remnants face similar issues, suggesting that quantum gravity should not possess global symmetries \cite{Susskind:1995da}.

With the above discussion, for very large $M$ with $\Delta\zeta_{Q_2}\sim\,0$ and $\xi\sim\,2$, we have
\begin{eqnarray}
&&\frac{1}{T_ {s}}\frac{dS^{\rm total}_{NG}}{dt}\sim\frac{M}{2}\label{complexiy7}
\end{eqnarray}
Our complex action conjecture (\ref{complexiy7}) is validated by the fact that black holes saturate Lloyd's bound \cite{Brown:2015bva, Lloyd:2000cry}. The "Complexity=Action" (CA) duality in Horndeski gravity offers a compelling framework for conceptualizing black holes as nature's fastest computers \cite{Lloyd:2000cry}. Furthermore, equation (\ref{complexiy7}) serves as a bridge between the mathematical formulation of complexity and its physical interpretation in the context of black hole thermodynamics and quantum gravity. By incorporating the effects of angular momentum, boundary conditions, and UV-IR dynamics, it provides a comprehensive framework for understanding how complexity evolves in holographic systems. This not only validates the CA conjecture but also opens new avenues for exploring the fundamental nature of spacetime, information, and computation in the universe. In the context of the CA duality, while discussing the non-radiation of charge during complete Hawking evaporation, we find that remnants of information persist. This information is encoded by
\begin{eqnarray}
\frac{1}{T_ {s}}\frac{dS^{\rm total}_{NG}}{dt}=-\frac{\Delta\zeta_{Q_2}}{2\sqrt{-\alpha/\gamma}}\left(1-\frac{\xi}{4}\right)\ln(2)\label{complexiy8}
\end{eqnarray}

As the black hole decays, the residual information corrections obtained through  "$Complexity=Action$" (CA) conjecture indicate that remnants are small objects approximately the size and mass of the Planck scale. In this context, a finite value of Newton's constant $G_{N}$ governs the gravitational force at asymptotic distances \cite{Susskind:1995da}.

\section{Conclusions}\label{v6}
The connection between the growth of holographic complexity and the minimal entropy of a rotating BTZ black hole enhances our understanding of the interplay between quantum effects and string behavior in gravitational backgrounds, as explored through the AdS/BCFT correspondence. Our study highlights how quantum aspects of the worldsheet of a probe string significantly contribute to quantum field theory in curved spacetime, with profound implications for fundamental physics. This residual information offers valuable insights into the complex dynamics of quantum systems within black hole geometries, advancing our understanding of quantum gravity in such environments.

Our investigation into holographic complexity via the AdS/BCFT correspondence reveals that corrections in the probe string worldsheet within the BTZ black hole scenario contribute to the complexity's growth. Additionally, as discussed in \cite{Zhou:2024pbb}, quantum fluctuations in the probe string worldsheet enhance our understanding of the relationship between complexity growth and correlation functions. These corrections, derived through the AdS/BCFT duality framework, provide valuable insights into the dynamics of quantum systems in complex gravitational environments.

The quantum complexity of a holographic state can be associated with the action of a Wheeler-DeWitt patch within a specific spacetime region. This association suggests a novel approach to understanding complexity, where the action of the spatial region serves as a promising and precise measure. As demonstrated by \cite{Brown:2015lvg}, this conjecture holds across various black hole configurations. In our study, we extend its validity to the Horndeski scenario, offering new perspectives for exploring the computational capabilities of black holes in the realm of quantum complexity.

The concept of residual information as boundary minimal entropy suggests the existence of a new ultraviolet scale, significantly below the Planck scale, where new physics emerges, potentially near the Grand Unified Theory (GUT) scale \cite{Arkani-Hamed:2006emk}. This conjecture implies that light elementary electric and magnetic objects must adhere to specific mass-to-charge ratios. Our findings support Arkani-Hamed's \cite{Arkani-Hamed:2006emk} argument that a universal limit on the strength of gravity relative to gauge forces provides fresh insights into string theory, black holes, and the fundamental nature of gravity as the weakest force.


\section*{Authors' Contributions}
All authors contributed to the study conception and design. Material preparation, data collection and analysis were performed by Fabiano F. Santos. The first draft of the manuscript was written by  Fabiano F. Santos commented on previous versions of the manuscript. All authors read and approved the final manuscript.

\section*{Data Availability}
\begin{enumerate}
\item[$\bullet$] The data are in he repository https://arxiv.org/abs/2407.10004
\item[$\bullet$] All original data for this work can be found at https://arxiv.org/abs/2407.10004
\end{enumerate}

\section*{Conflicts of Interest}
The authors declare that there is no conflict of interest.

\section*{Ethical Considerations}
The authors have diligently addressed ethical concerns, such as informed consent, plagiarism, data fabrication, misconduct, falsification, double publication, redundancy, submission, and other related matters.

\section*{Funding}
This research did not receive any grant from funding agencies in the public, commercial, or non-profit sectors.


\section*{Acknowledgment}
Would like to thank Henrique Boschi-Filho for fruitful discussions.


%
%
%
%

\end{document}